\documentclass[aip,jcp,reprint,amssymb,amsmath,superscriptaddress,groupedaddress,frontmatterverbose,]{revtex4-2}

\usepackage[title]{appendix}
\usepackage{graphicx, tikz, orcidlink}
\usepackage{epstopdf}
\usepackage{ascmac}
\usepackage{ulem}
\usepackage{cancel}
\usepackage{here}
\usepackage{physics}
\usepackage{bm}
\usepackage{comment}
\usepackage{booktabs}

\DeclareGraphicsExtensions{{.eps,.pdf,.png}}
\graphicspath{{./}}

\begin{document}

\title{A Multimode Classical Hierarchical Fokker–Planck Equations Approach to Molecular Vibrations: Simulating Two-Dimensional Spectra}
\date{Last updated: \today}

\author{Ryotaro Hoshino\orcidlink{0009-0004-4208-326X}}\email[Author to whom correspondence should be addressed: ]{hoshino.ryotaro.75x@st.kyoto-u.ac.jp}
\author{Yoshitaka Tanimura\orcidlink{0000-0002-7913-054X}}
\email[Author to whom correspondence should be addressed: ]{tanimura.yoshitaka.5w@kyoto-u.jp}
\affiliation{Department of Chemistry, Graduate School of Science,
Kyoto University, Kyoto 606-8502, Japan}

\begin{abstract}
The multimode Brownian model with nonlinear system-bath coupling offers a flexible framework for studying both intra- and intermolecular vibrational modes in condensed-phase molecular systems. This approach allows us to calculate linear and nonlinear spectra of molecular vibrations and to examine thermal effects—such as anharmonicity, energy relaxation, and dephasing—as reflected in the spectral peak profiles. In this study, we present a computer program based on classical hierarchical Fokker-Planck equations applied to three vibrational modes of a molecular liquid. The primary objective of developing this code was to simulate the two-dimensional correlation spectrum of the intramolecular modes of liquid water. [R. Hoshino and Y. Tanimura, J. Chem. Phys. 162, 044105 (2025)]. 
The code has been further refined to optimize grid selection and numerical integration routines for graphics processing units (GPUs). As a demonstration, we apply this setup to simulate three interacting modes representing intermolecular vibrations in water, and calculate the resulting two-dimensional terahertz-Raman signals. The code and example routines are available in the supplementary material.
\end{abstract}

\maketitle

\section{INTRODUCTION}
Solvents serve as the ideal medium for facilitating chemical reactions.\cite{covington_physical_chemistry_1978,mabesoone_solute_2022} The dynamic behavior of solvents, in particular water, involves not only thermal phenomena such as fluctuations that supply activation energy to reactions and dissipation that absorbs heat (enthalpy) of reaction, but also extremely fast processes such as phase relaxation, hydrogen bond rearrangement, and proton transfer, making their analysis extremely difficult.\cite{Ohmine_ChemRev93,OCSACR1999,Nibbering2004UltrafastVD,bagchi_2013} Experimentally, the method for analyzing such fast and complex solvent processes is ultrafast nonlinear laser spectroscopy, particularly coherent two-dimensional (2D) spectroscopy.\cite{TM93JCP,mukamel1999principles,Cho2009,Hamm2011ConceptsAM,HammPerspH2O2017}
Its distinctive feature is its ability to detect anharmonicity of vibrational peaks, including mode-mode coupling, and to quantitatively evaluate the time scale of vibrational dephasing, which is the cause of inhomogeneous broadening.\cite{ElsaesserH2O,ElsaesserDwaynePNAS2008,TokmakoffNat2013,Tokmakoff2015,Tokmakoff2016H2O,Tokmakoff2022,HammTHz2012,Hamm2013PNAS,hamm2014,Hamm20252DTHz,Blake20162DThzRaman,Blake20172DThzRaman,Blake20192DThzRaman,Blake20202DThzRaman,
grechko2018,Bonn2DTZIFvis2021,Begusic2023}  Such 2D profiles with a nonlinear response function as the experimental observable are highly sensitive, and theoretical backup is essential to fully utilize its advantages.\cite{TI09ACR,JansenSkiner2010,JansenChoShinji2DVPerspe2019}

Full molecular dynamics (MD) simulations, which mimic experimental procedures to generate 2D spectra, offer a promising methodology for extracting detailed information such as signal intensities across various solvents.\cite{Shinji2DRaman2006,HT06JCP,LHDHT08JCP,HT08JCP,YagasakiSaitoJCP20082DIR,Saito1995water3modes,Saito1997water3modes,YagasakiSaitoJCP2011Relax,Yagasaki_ARPC64,Wei2015Nagata2DRamanTHz,IHT14JCP,JIT16CP,IHT16JPCL,IIT15JCP,ImotXanteasSaitoJCP2013H2O,Imotobend-lib2015}  However, 2D spectral features are highly sensitive to molecular properties, and accurate reproduction requires quantum-level treatment of both nuclear motion and electronic structure.
 \cite{ST11JPCA,TT23JCP1,TT23JCP2,HT25JCP1} Despite recent advances, the computational cost remains prohibitive, and MD-based approaches still fall short of fully replicating experimental observations. 
 Moreover, even when spectral reproduction is successful, interpretation remains challenging due to the complex dependence of MD results on all signal-generating factors, including dipolar interactions.
To address these limitations, quantum-implemented MD simulations have been developed, incorporating stochastic dynamics\cite{SkinnerStochs2003} and excitonic wave-function based frameworks.\cite{Mukamel2009,JansenSkiner2010}

Although the multimode Brownian oscillator (BO) model is not derived from first principles,\cite{TM93JCP,mukamel1999principles}  it provides a practical framework for describing inter- and intramolecular vibrations as principal harmonic oscillator modes.
It also accounts for fluctuations and dissipation resulting from interactions with a thermal bath, making it an effective tool for characterizing molecular systems in condensed phases.
While multimode Brownian model enables the analytical solution of the spectrum when the principal modes are harmonic,\cite{TM93JCP}  it is necessary to consider the anharmonicity of the modes,\cite{T98CP,ST11JPCA}  as well as the anharmonic coupling between the modes\cite{TI09ACR} to predict the profile of 2D spectral peaks,

Moreover, to incorporate the effects of vibrational dephasing, it is necessary to consider the nonlinear, non-perturbative, and non-Markovian system-enviroment interactions.\cite{T06JPSJ,T20JCP} Although computationally expensive, nonlinear spectra for such model systems can be evaluated numerically accurately using hierarchical Fokker-Planck equations (HFPE), both in classical (CHFPE) and quantum (QHFPE) cases.\cite{ST11JPCA,HT25JCP1}

This dual approach to Hamiltonian systems—bridging classical and quantum mechanics—offers powerful analytical insights.\cite{ST11JPCA} By aligning parameters to reproduce classical MD results through CHFPE,\cite{IT16JCP}  and then leveraging quantum mechanical solutions using QHFPE,\cite{TT23JCP1,TT23JCP2} we can navigate deeper into the actual quantum dynamics.
Furthermore, classical calculations are regarded as adequate for simulating 2D intermolecular vibrational spectroscopy, including 2D Raman spectroscopy\cite{Shinji2DRaman2006,HT06JCP,LHDHT08JCP}  and 2D THz-Raman spectroscopy.\cite{IHT14JCP,IIT15JCP,JIT16CP,IHT16JPCL} 
This is because quantum coherence is destroyed due to the thermal excitation energy being close to the vibrational excitation energy.

Quantum calculations of the 2D IR spectrum of water based on the QHFPE framework have thus far been limited to two-mode Brownian systems, typically involving the stretching and bending vibrational modes.\cite{TT23JCP1,TT23JCP2} 
Although a two-mode model is sufficient for accounting for mode coupling mechanisms, three modes are necessary for describing energy and coherent transfer processes.
Furthermore, since 2D spectroscopy possesses the capability to experimentally explore this, three-mode research becomes crucial.
Recently, we conducted classical simulations for three-mode Brownian systems, which include symmetric, antisymmetric, and bending modes.\cite{HT25JCP1} While classical calculations cannot adequately capture ultrafast relaxation mediated by the coherence of intramolecular modes,\cite{ST11JPCA,HT25JCP1} they are well-suited for modeling slower energy transitions between modes. This approach is particularly applicable to 2D THz-Raman spectroscopy, which focuses on intermolecular modes.

In this paper, we present a computational software developed for simulating 2D IR spectra based on CHFPE framework, employing an efficient numerical algorithm accelerated by a graphics processing unit (GPU).\cite{HT25JCP1} We demonstrate this by computing 2D THz-Raman signals including three intermolecular vibration modes.
In Sec.~\ref{sec:model}, we introduce the model Hamiltonian, CHFPE, and optical observables to be simulated.
In Sec.~\ref{NIC}, we explain algorithms for developing efficient code for CHFPE.
In Sec.~\ref{CHFPEcode} flowchart of the code is given.  In Sec. \ref{NumericalDemo}  we demonstrate the capability of our codes by simulating 2D THz-Raman spectra.
Our concluding remarks are presented in Sec.~\ref{sec:conclude}. The C++ source code for CHFPE are provided in the supplementary material.

\section{Formulation}
\label{sec:model}
\subsection{Multimode anharmonic Brownian model}\label{sub:MMBO}
We consider a model consisting of three primary intermolecular and/or intramolecular modes.  
These modes are described by dimensionless vibrational coordinates $\bm{q}=(q_1, q_{2}, q_3)$. Each mode is independently coupled to the other optically inactive modes, which constitute a bath system represented by an ensemble of harmonic oscillators. The total Hamiltonian can then be expressed as\cite{IT16JCP,ST11JPCA,T98CP,T06JPSJ,T20JCP,TS20JPSJ,KT04JCP,IIT15JCP,TT23JCP1,TT23JCP2,HT25JCP1,UT20JCTC} 
\begin{align}
\hat{H}_{tot}=  \sum_{s} \qty( \hat{H}_{A}^{(s)} +\hat{H}_{I}^{(s)} +\hat{H}_{B}^{(s)}) +  \sum_{s<s'} \hat{U}_{ss'}\qty(\hat{q}_s, \hat{q}_{s'}),
\label{sec:Total Hamiltonian}
\end{align}
where
\begin{align}
\hat{H}_{A}^{(s)}= \frac{\hat{p}_s^{2}}{2m_s} +\hat U_s(\hat{q}_s)
\label{sec:System Hamiltonian}
\end{align}
is the Hamiltonian for the $s$th mode, with mass $m_s$, coordinate ${\hat{q}_s}$, and momentum ${\hat p_s}$; and
\begin{align}
\hat U_s(\hat{q}_s)= \frac{1}{2} m_s \nu_s^2 \hat{q}_s^2 +\frac{1}{3!}g_{s^3}q_{s}^3
\label{sec: Potenentials}
\end{align}
is the anharmonic potential for the $s$th mode, described by the frequency  $\nu_s$ and cubic anharmonicity $g_{s^3}$. 
The anhramonic coupling between the $s$th and $s'$th modes is given by 
\begin{align}
\hat{U}_{ss'}(\hat{q}_s, \hat{q}_{s'}) = g_{s{s'}}\hat{q}_s\hat{q}_{s'} + \frac{1}{6}  \qty(g_{s^2s'}\hat{q}_s^2 \hat{q}_{s'} + g_{s{s'}^2} \hat{q}_s \hat{q}_{s'}^2 ),
\label{sec: Potential ss'}
\end{align}
where $g_{s{s'}}$ represents the second-order harmonicity, and $g_{s^2s'}$ and $g_{s{s'}^2}$ represent the third-order anharmonicity.  The bath Hamiltonian for the $s$th mode is expressed as
\begin{align}
\hat{H}_{B}^{(s)}= \sum_{j_s}\qty[\frac{\hat{p}_{j_s}^{2}}{2m_{j_s}}+\frac{m_{j_s}\omega_{j_s}^{2}}{2}\qty(\hat{x}_{j_s}-\alpha_{j_s} \hat{V}_s(\hat{ q}_s) )^2],
\label{sec: Bath Hamiltonian}
\end{align}
where the momentum, coordinate, mass, and
frequency of the $j_s$th bath oscillator are given by ${p}_{j_s}$, ${x}_{j_{s}}$, $m_{j_{s}}$ and
$\omega _{{j_s}}$, respectively.  The system-bath interaction, defined as
\begin{align}
  {H}^{(s)}_{\mathrm{I}}&=- V_{s}({q_s})\sum _{j_s}\alpha _{j_s}{x}_{j_s},
  \label{eq:h_int}
\end{align}
consists of linear-linear (LL)\cite{TM93JCP,T98CP}
and square-linear (SL) system-bath interactions,\cite{TS20JPSJ,KT04JCP,UT20JCTC,OT97PRE}
$V_{s}({q_s})\equiv V^{(s)}_{\mathrm{LL}}{q_s}+V^{(s)}_{\mathrm{SL}}{q_s}^{2}/2$, with coupling strengths $V^{(s)}_{\mathrm{LL}}$, $V^{(s)}_{\mathrm{SL}}$,
and $\alpha _{j_s}$.\cite{IT16JCP,ST11JPCA,T06JPSJ,T20JCP,TS20JPSJ,KT04JCP,IIT15JCP,TT23JCP1,TT23JCP2,HT25JCP1,T98CP} 
While the LL interaction mainly contributes to energy relaxation, the LL+SL system--bath interaction causes vibration dephasing in the case of slow modulation, owing to frequency fluctuations in the system oscillations. \cite{TS20JPSJ,T06JPSJ,T20JCP,IT16JCP} 
The bath property is then characterized by the Drude Spectral Distribution Function (SDF) expressed as\cite{T06JPSJ,TI09ACR} 
\begin{equation}
  J_s(\omega)=\frac{m_s \zeta_s}{2\pi}\frac{ \gamma_{s}^{2}\omega}{\omega^{2}+\gamma_s^{2}}, 
\label{eq:drude}
\end{equation}
where  $\zeta_s$ is the system--bath coupling strength, and  $\gamma_s$ represents
the width of the SDF for mode $s$, which relates to the
vibrational dephasing time, defined as  $\tau_s$ = ${1}/{\gamma_s}$.

\subsection{Multimode CHFPE}
\label{sec:CHFPE}
The LL+SL anharmonic Brownian model has been studied by means of the HEOM formalism.\cite{T06JPSJ,T20JCP} In the classical case, the HEOM in the phase space for the system described by Eqs.~\eqref{sec:Total Hamiltonian}$-$\eqref{eq:drude} are expressed as\cite{IT16JCP,HT25JCP1}
\begin{align}
\label{eq:HEOM}
  \frac{ \partial{W^{(\bm{n})}(\bm{q}, \bm{p}; t)}}{\partial t}& = 
	 \left(\hat{L}(\bm{q}, \bm{p}) -\sum_{k} \sum_{s} n_{k}^{(s)} \gamma_{k}^{(s)} \right) W^{(\bm{n})}(\bm{q}, \bm{p}; t) \nonumber \\
	& +\sum_{k} \sum_{s}\hat{\Phi}_{k}^{(s)} W^{(\bm{n}+\bm{e}_{k}^{(s)})}(\bm{q}, \bm{p}; t)\nonumber \\
	&+\sum_{k} \sum_{s}\hat{\Theta}_{k}^{(s)} W^{(\bm{n}-\bm{e}_{k}^{(s)})}(\bm{q}, \bm{p}; t),
\end{align}
where $W^{(\bm{n})}(\bm{q},  \bm{p}; t) \equiv W^{(\bm{n})}(\{q_s\}, \{p_s\}; t)$ is the Wigner distribution function (WDF) as the function of the set of $\{q_s\}$ and $\{p_s\}$.  As we are considering the case of three modes, the hierarchical elements are expressed here as $\bm{n}$ = $(n_1,n_2,n_3)$, where each $s$th mode element is denoted by a positive integer $n_s$, and $\bm{e}_{s}$ is the unit vector for the $s$th space.

The classical Liouvillian $\hat{L}$ for the system Hamiltonian $H_{\rm sys}(\bm{q}, \bm{p}) \equiv \sum_{s} {H}_{A}^{(s)} + \sum_{s<s'} U_{ss'}\qty(\hat{q}_s, \hat{q}_{s'})$ can be expressed as
\begin{align}
\label{eq:CL_liouville}
\hat{L}(\bm{q}, \bm{p})  W(\bm{q}, \bm{p}) &\equiv \{ H_{\rm sys}(\bm{q}, \bm{p}) , W(\bm{q}, \bm{p})  \}_{\mathrm{PB}}, 
\end{align}
where $\{ \hspace{2mm},\hspace{2mm} \}_{\mathrm{PB}}$ is the Poisson bracket defined as
\begin{align}
\{ A, B \}_{\mathrm{PB}}  \equiv \sum_s \left( \frac{\partial A}{\partial q_s}\frac{\partial B}{\partial p_s} - \frac{\partial A}{\partial p_s}\frac{\partial B}{\partial q_s} \right)
\end{align}
for any functions $A$ and $B$.  Operators $\hat{\Phi}_{s}$ and $\hat{\Theta}_{s}$ represent the  energy exchange between the $s$th mode and the $s$th bath through fluctuation and dissipation, respectively. They are expressed as\cite{TS20JPSJ,KT04JCP,IT16JCP,TT23JCP1,TT23JCP2,HT25JCP1} 
\begin{align}
	\label{eq:Phi}
	\hat{\Phi}_{s}=\frac{ \partial V_s(q_s)}{\partial q_s} \frac{ \partial }{\partial p_s}
\end{align}
and
\begin{align}
	\label{eq:Theta}
	\hat{\Theta}_{s}=\frac{ m_s \zeta_s \gamma_s}{\beta}\frac{ \partial V_s(q_s)}{\partial q_s} \frac{ \partial }{\partial p_s}
        +\zeta_s \gamma_s p_s \frac{ \partial V_s(q_s)}{\partial q_s}, 
\end{align}
where $\zeta_s$ is the coupling strength, $\gamma_s$ is the inverse correlation time, and $\beta$ is the inverse temperature.

\subsection{Linear and non-Linear spectra}
\label{sec:spectruman}

We now consider the optical measurements where the molecular system is interacting with a laser field, $E(t)$. 
The nonlinear elements of polarizability and dipole are essential to 2D spectroscopy.  Here we assume\cite{ST11JPCA,T98CP, IT16JCP,TT23JCP1,TT23JCP2,HT25JCP1}
\begin{eqnarray}
{\mu}({\boldsymbol q}) = \sum_s\mu_sq_s + \frac{1}{2!}\sum_{s,s'}\mu_{ss'}q_sq_{s'}
 \label{NLdip}
\end{eqnarray}
and
\begin{eqnarray}
{\Pi}({\boldsymbol q}) = \sum_s\Pi_sq_s + \frac{1}{2!}\sum_{s,s'}\Pi_{ss'}q_sq_{s'}\label{NLpol},
\end{eqnarray}
where $\mu_s$ and $\mu_{ss'}$ are the linear and nonlinear elements of the dipole moment, and $\Pi_s$, and $\Pi_{ss'}$ are those of the polarizability, respectively. 
For infrared (IR) and Raman spectroscopies, the laser interaction is then expressed as $H_{\rm IR}=- E(t){\mu}({\boldsymbol q})$ and  $H_{\rm Raman}=- E^2(t){\Pi}({\boldsymbol q})$, respectively. 

In the Wigner representation, the first, second, and third-order response functions are expressed as\cite{IT16JCP,T06JPSJ,T20JCP,TS20JPSJ,KT04JCP} 
\begin{equation}
R^{(1)}(t_{1})
= \iint \mathrm{d}\bm{p}\,\mathrm{d}\bm{q}\;
A(\bm{q})\,\mathcal{G}(t_{1})
A^{\times}(\bm{q})
W^{\mathrm{eq}}(\bm{p},\bm{q}),
\label{1Dresponse}
\end{equation}
\begin{eqnarray}
R^{(2)}(t_{2},t_{1})
&&= \iint \mathrm{d}\bm{p}\,\mathrm{d}\bm{q}\;
C(\bm{q})\,\mathcal{G}(t_{2})  \nonumber \\
&&\times B^{\times}(\bm{q})
\mathcal{G}(t_{1})
A^{\times}(\bm{q})
W^{\mathrm{eq}}(\bm{p},\bm{q}),
\label{2Dresponse}
\end{eqnarray}
and
\begin{eqnarray}
R^{(3)}(t_{3},t_{2},t_{1}) &&
= \iint \mathrm{d}\bm{p}\,\mathrm{d}\bm{q}\;
D(\bm{q})\,\mathcal{G}(t_{3})
C^{\times}(\bm{q})
\mathcal{G}(t_{2}) \nonumber \\
&&\times
B^{\times}(\bm{q})
\mathcal{G}(t_{1})
A^{\times}(\bm{q})
W^{\mathrm{eq}}(\bm{p},\bm{q}),
\label{3Dresponse}
\end{eqnarray}
where $\mathcal{G}(t)$ is the Green's function for the system+bath Hamiltonian without laser interaction described as CHFPE and  
 $W^{\mathrm{eq}}(p, q)$ is the equilibrium distribution described in terms of the CHFPE elements.\cite{IT16JCP,T06JPSJ,T20JCP,TS20JPSJ,KT04JCP,HT25JCP1,T98CP} 
The optical operators $A(\bm{q})$, $B(\bm{q})$, $C(\bm{q})$, and $D(\bm{q})$ are either the Raman polarizability function 
${\Pi}({\boldsymbol q})$ or dipole function ${\mu}({\boldsymbol q})$, depending upon the measurements. For any function $X(\bm{q})$ and $W(\bm{p},\bm{q})$, the operator $X^{\times}(\bm{q})$ in the classical case is defined as
\begin{equation}
X^{\times}(\bm{q}) W(\bm{p},\bm{q}) = \sum_s  \frac{\partial X(\bm{q})}{\partial q_s}\frac{\partial W}{\partial p_s}.
\label{Opticalexc}
\end{equation}
The above expressions enable us to calculate the response functions using the equations of motion, and 
give us an intuitive picture of measurements in higher-order optical processes.\cite{ST11JPCA,IT16JCP,T06JPSJ,T20JCP,TS20JPSJ,KT04JCP,IIT15JCP,IT16JCP,TT23JCP1,TT23JCP2,HT25JCP1,T98CP}

The linear absorption spectrum or Raman spectrum, 2D IR-Raman spectra,\cite{IT16JCP,TT23JCP1,grechko2018,Bonn2DTZIFvis2021,Begusic2023} and the third-order IR spectrum are then given by\cite{Cho2009,Hamm2011ConceptsAM,HammPerspH2O2017,TI09ACR,TT23JCP2,HT25JCP1,
ST11JPCA,IT16JCP,T06JPSJ,T20JCP,TS20JPSJ,KT04JCP,IIT15JCP,IT16JCP,TT23JCP1,TT23JCP2,HT25JCP1,T98CP}
\begin{equation}
I(\omega_{1})
= \omega_{1} \Im \left\{
\int_{0}^{\infty} \mathrm{d}t_{1}\,
e^{i\omega_{1}t_{1}}
\,R^{(1)}(t_{1})
\right\},
\label{1DIR_Raman}
\end{equation}
\begin{align}
I(\omega_{1},&\omega_{2})
= \nonumber \\
& \Im \left\{
\int_{0}^{\infty}\mathrm{d}t_{1}
\int_{0}^{\infty}\mathrm{d}t_{2}\,
e^{-i\omega_{1}t_{1}}
e^{-i\omega_{2}t_{2}}
\,R^{(2)}(t_{2},t_{1})
\right\},
\label{2DThz_Raman}
\end{align}
and
\begin{align}
\label{noncorrelation}
I(\omega_{3},&t_{2},\omega_{1})
= \nonumber \\
&\Im \Biggl\{
\int_{0}^{\infty}\!\mathrm{d}t_{1}
\int_{0}^{\infty}\!\mathrm{d}t_{3}\,
e^{-i\omega_{1}t_{1}}
e^{i\omega_{3}t_{3}}
\,R^{(3)}(t_{3},t_{2},t_{1})
\Biggr\},
\nonumber\\
\end{align}
respectively.

\section{NUMERICAL IMPLEMENTATION OF CHFPE}
\label{NIC}

\subsection{Momentum space representation using Hermite polynomial}
The CHFPE, Eq.~\eqref{eq:HEOM}, converge very slowly as difference equations with a discrete mesh in phase space.  
As in the case of the Kramers equation, in which the relaxation operator (the Ornstein-Uhlenbeck operator) can be expressed in terms of Hermite polynomials, the CHFPE can also be efficiently expanded using these polynomials.\cite{IIT15JCP,IT16JCP} In the three modes case, the CHFPE for the set of functions $c^{(\bm{n})}_{k, l,m}({\boldsymbol q};t)$ $\equiv$ $c^{(n_1, n_2, n_3)}_{k, l,m}(q_1, q_2,q_3;t)$ are expressed as
\begin{eqnarray}
& W^{(\bm{n})}({\boldsymbol p},{\boldsymbol q};t) = \psi_0(p_{1})\psi_0(p_{2})\psi_0(p_{3})e^{-{\beta}U({\boldsymbol q})/2
} \nonumber \\
&\sum_{k=0}^{\infty}\sum_{l=0}^{\infty}\sum_{m=0}^{\infty}c^{(\bm{n})}_{k, l ,m}({\boldsymbol q};t)\psi_k(p_{1})\psi_l(p_{2})\psi_m(p_{3}), \nonumber \\
\end{eqnarray}
where $U({\boldsymbol q})$ $\equiv$ $\sum_{s=1}^{3}U_{s}(q_{s}) + \sum_{s=1}^{3}\sum_{s'=s+1}^{3} U_{ss'}(q_{s}, q_{s'})$ is the potential of the system and $\psi_j(p_{s})$ is the $j$th Hermite function,
\begin{eqnarray}
 \psi_j(p_{s}) = \frac{1}{\sqrt{2^jj!a_s\sqrt{\pi}}}H_j\left(\frac{p_s}{a_s}\right){\exp}\left(-\frac{p_s^2}{2a_s^2}\right),
\end{eqnarray}
with $H_j(x)$ the $j$th Hermite polynomial and $a_s = \sqrt{2m_sk_BT}$.

The equations of motion for $c^{(\bm{n})}_{k, l,m}({\boldsymbol q};t)$ are presented in Appendix \ref{Sec:HEOMDIS}

\subsection{Compact Finite Difference Scheme (CFDS)}
A compact finite difference scheme is a numerical method that approximates derivatives with high accuracy while minimizing stencil size. Here we employ the algorithm based on Pad\'e-type compact finite differences.\cite{LELE199216}
We discretized  $c^{(\bm{n})}_{k, l,m}({\boldsymbol q};t)$ and define the vector $\bm{c}$, whose elements $\{ c_i \}$ correspond to a discrete set of positions $\{ q_s \}$, as follows:
\begin{eqnarray}
{c_{k,l,m}^{(\bm n )}} \in \mathbb{R}^{N_1 \times N_2 \times N_3}
\quad \Rightarrow \quad
\bm{c} \in \mathbb{R}^{N_1 N_2 N_3}.
\end{eqnarray}
The first derivative of $\bm{c}$ at the nodal (mesh) point $i$ is represented as $c'_i$. Using a three-point tridiagonal stencil, the derivatives of $c$ can be computed with an accuracy up to sixth order, as detailed below:
\begin{eqnarray}
\left( \mathbf{A_1} \otimes \mathbf{I}_{N_2} \otimes \mathbf{I}_{N_3} \right)
\frac{\partial \bm{c}}{\partial q_1}
=
\left( \mathbf{B_1} \otimes \mathbf{I}_{N_2} \otimes \mathbf{I}_{N_3} \right)
\bm{c},
\end{eqnarray}
\begin{eqnarray}
\left( \mathbf{I}_{N_1} \otimes \mathbf{A_2} \otimes \mathbf{I}_{N_3} \right)
\frac{\partial \bm{c}}{\partial q_2}
=
\left( \mathbf{I}_{N_1} \otimes \mathbf{B_2} \otimes \mathbf{I}_{N_3} \right)
\bm{c},
\end{eqnarray}
and
\begin{eqnarray}
\left( \mathbf{I}_{N_1} \otimes \mathbf{I}_{N_2} \otimes \mathbf{A_3} \right)
\frac{\partial \bm{c}}{\partial q_3}
=
\left(  \mathbf{I}_{N_1} \otimes \mathbf{I}_{N_2} \otimes \mathbf{B_3} \right)
\bm{c},
\end{eqnarray}
where the matrices are defined as
\begin{eqnarray}
\mathbf{A}_i =
\begin{bmatrix}
1 & \alpha & 0 & \cdots & 0 \\
\alpha & 1 & \alpha & \ddots & \vdots \\
0 & \alpha & 1 & \ddots & 0 \\
\vdots & \ddots & \ddots & \ddots & \alpha \\
0 & \cdots & 0 & \alpha & 1
\end{bmatrix},
\end{eqnarray}
and
\begin{eqnarray}
\mathbf{B}_i = 
\begin{bmatrix}
0 & a & b & 0 & \cdots & 0 \\
-a & 0 & a & b & \ddots & \vdots \\
-b & -a & 0 & a & b & 0 \\
\vdots & \ddots & \ddots & \ddots & \ddots & \vdots \\
0 & \cdots & -b & -a & 0 & a \\
0 & \cdots & 0 & -b & -a & 0
\end{bmatrix}
\end{eqnarray}
with the constants
\begin{eqnarray}
\alpha = \frac{1}{3}, \beta = \frac{7}{9}, \gamma = \frac{1}{9}, a = \frac{\beta}{2} = \frac{7}{18}, 
\end{eqnarray}
and
\begin{eqnarray}
b = \frac{\gamma}{4} = \frac{1}{36}.
\end{eqnarray}

\subsection{Parallel computing method for tridiagonal simultaneous differential equations}
We used the Parallel Cyclic Reduction (PCR) method,\cite{Louie1985PCR}  which is suitable for high-speed parallel processing, to solve the above triple-diagonal simultaneous equations.

In the PCR method, each process performs operations independently while reducing the system size by half.  Then the calculation ``($\text{reduction}$) $\rightarrow$ ($\text{solution}$) $\rightarrow$  ($\text{back-substitution}$)'' is performed recursively in parallel on the time scale $\mathcal{O}(\log N)$. This implementation was utilized with Compute Unified Device Architecture (CUDA), which performs parallelization for the Graphics Processing Unit (GPU).

\subsection{Non-Uniform Mesh in coordinate space}

A non-uniform mesh was implemented across all $q_s$ directions: finer meshes were employed in regions with steep gradients in physical quantities to maximize the benefits of the higher-order finite difference method.
Conversely, a coarse mesh is used in regions where the gradient of physical quantities is moderate, thereby reducing grid numbers and lowering computational and communication costs.
Our software supports three types of non-uniform mesh specification: (i) user-defined meshes, (ii) automatically optimized meshes via routines employing the Optuna library, and (iii) meshes parsed directly from configuration data.

\section{CHFPE code for simulating multi-dimensional spectra}
\label{CHFPEcode}
\begin{figure*}[!t]
\centering
\includegraphics[width=180mm]{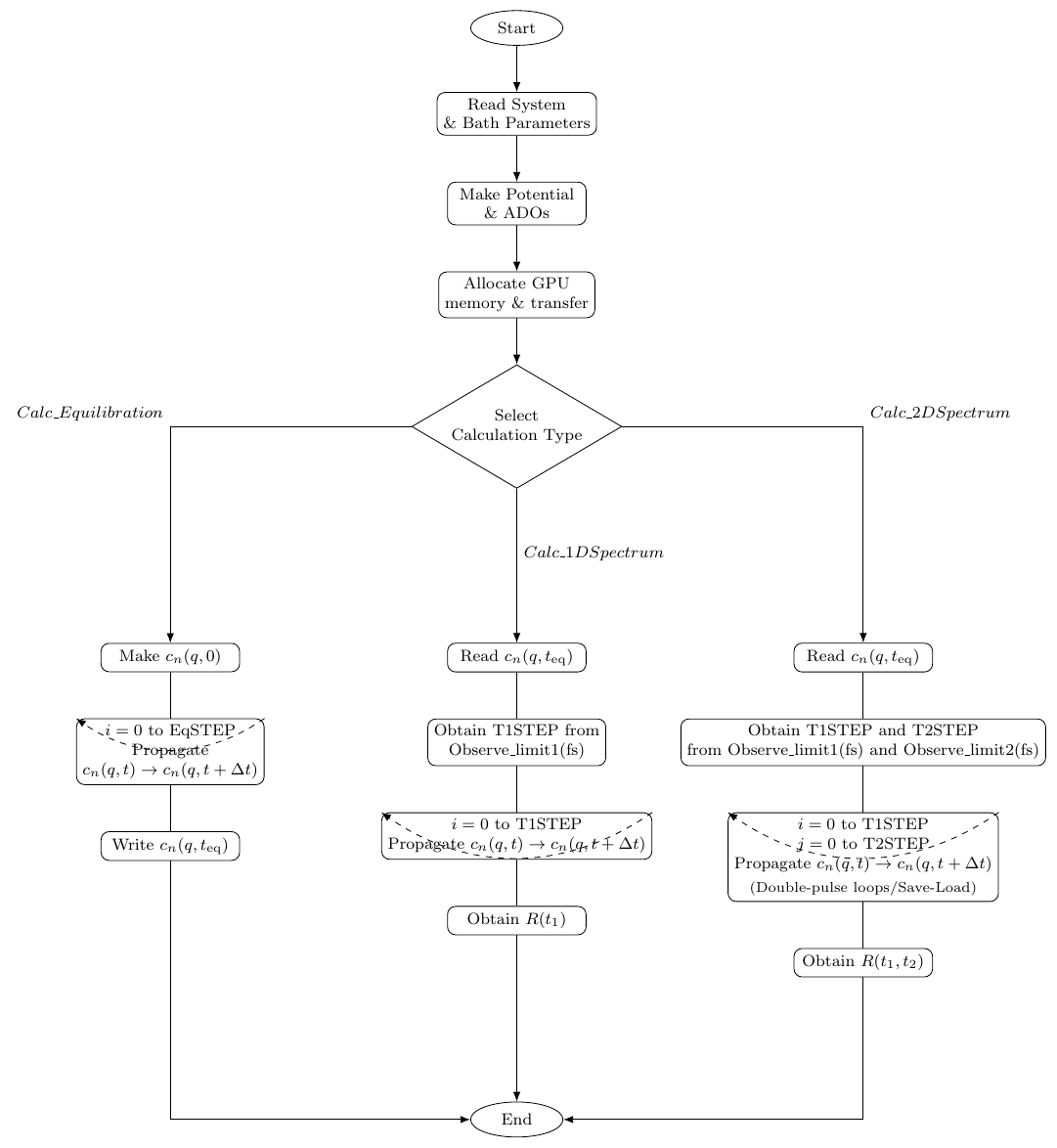}
\caption{\label{algo}
Flowchart of the CHFPE code for computing 1D and 2D spectra
}
\end{figure*}

We have developed a computer code to efficiently compute multi-dimensional response functions and spectra using CHFPE, employing the algorithm described above.  The process performed by the computer program consists of three steps: $\bf A$ Initialization, $\bf B$ equilibrium state calculation, and $\bf C$ calculating response functions and spectra. A flowchart of the code is presented in Fig. \ref{algo}

\subsection{Initialization}
The input file specifies the system and heat bath parameter variables needed for calculating 1D and 2D spectra. For the present demonstration, these variables have been chosen to replicate the 1D and 2D spectra obtained from MD simulations of water.\cite{IHT14JCP,IT16JCP} 
The file also defines working variables, such as the number of Auxiliary Density Operator (ADO) of CHFPE and the time step for integration. 
Finally, the parallel computing environment is initialized. This process involves allocating GPU memory using the CUDA library and transferring the initial ADO array state to the GPU device.

\subsection{Equilibration}
After initialization is complete, calculations are performed to bring the total system to equilibrium. This is done by propagating the CHFPE over time: $W^{(\bm{n})}(\bm{q}, \bm{p}; 0)\to W^{(\bm{n})}(\bm{q}, \bm{p}; \Delta t)\to W^{(\bm{n})}(\bm{q}, \bm{p}; t_{\rm eq})$. 
This process continues until a steady state is reached after a sufficient time step (EqSTEP).  The final results with ADOs obtained after time evolution are saved in a file as the equilibrium distribution.
The following calculation employs this equilibrium state as the initial condition and computes the linear and nonlinear response functions.

\subsection{Calculations of Response Functions and Spectra}
Perform the following calculations according to the options selected in the input file.

\subsubsection{1D IR and Raman spectra}
Read the equilibrium distribution $W^{(\bm{n})}(\bm{q}, \bm{p}; t_{\rm eq})$ from the file. The pulse width, convergence condition, time step width, etc., are read from the spectroscopy setup file. 
After Raman and IR excitation, denoted as ${\Pi}^{\times}({\boldsymbol q})$ or ${\mu}^{\times}({\boldsymbol q})$ is applied, the time integration of the CHFPE is performed.  The Raman and IR responses $R_{\rm IR}(t_{1})$ and $R_{\rm Raman}(t_{1})$ are computed using Eq.\eqref{1Dresponse} for each $t_{1}$.
Finally, we evaluated the Raman and linear absorption spectra using Eq. \eqref{1DIR_Raman} via fast Fourier transformation (FFT), and the results are saved in a file.

\subsubsection{2D THz-Raman and IR-Raman spectra}
Compute the 2D spectrum defined by the second-order response function given in Eq. \eqref{2DThz_Raman} for different IR and Raman configurations, such as  $R_{\rm TRT}(t_{2},t_{1})$ for $A(\bm{q})={\mu}({\boldsymbol q})$. $B(\bm{q})={\Pi}({\boldsymbol q})$, and $C(\bm{q})={\mu}({\boldsymbol q})$. 
The time evolution of the CHFPE is performed as a double loop of $t_1$ and $t_2$, varying from 0 to $T_1$ and from 0 to $T_2$, respectively.   Store $R(t_2, t_1)$ at each time point, then use a 2D FFT to obtain $I(\omega_1, \omega_2)$, defined by Eq. \eqref{2DThz_Raman}.

\subsubsection{2D IR spectrum}
The third-order response function $R^{(3)}(t_3, t_2, t_1)$ is computed by evaluating the time evolution of the corresponding Liouville paths separately. We then obtain 2D  spectrum from Eq \eqref{noncorrelation}. In these calculations, temporary intermediate results are saved, and GPU memory is reused to achieve efficient computation of large data. Since the computation of 2D spectra is data-intensive, it is necessary to optimize memory management in the GPU and communication between the host and device.


\section{Numerical demonstrations: 2D THz-Raman spectra}
\label{NumericalDemo}

\begin{table*}[!tb]  
  \caption{\label{tab:FitAll1}The parameter values of the MAB model for the (3) hindered rotation (HR),   (4) high frequency translational (HT) motion, and (5) the low frequency translational (LT) motion that were chosen to reproduce the 1D and 2D vibrational spectra of liquid water obtained from the full MD simulations.\cite{IHT16JPCL}
We set the fundamental frequency and temperature as $\omega_{0} = 4000$ cm$^{-1}$, and $T = 300$ $K$.
The correlation times of the Gaussian-Markovian bath noise, $\tau_s=1/\gamma_s$, are also presented.}
\begin{tabular}{cccccccccccc}
    \hline \hline
    $s$  & $\omega_s$ (cm$^{-1}$) & $\gamma_s/\omega_0$ & $\tau_s$ (fs) & $\tilde{\zeta}_s$ & $\tilde{V}_{LL}^{(s)}$ & $\tilde{V}_{SL}^{(s)}$ & $\tilde{g}_{s^3}$ & $\tilde{\mu}_{s}$ & $\tilde{\mu}_{ss}$ & $\tilde{\Pi}_{s}$ & $\tilde{\Pi}_{ss}$ \\
    \hline
    $3$  & $600 $ & $0.075$ & $111$  & $5.90$ & $0$ & $1.0$ & $ 1.0{\times}10^{-1}$ & $2.0$ & $  0  $              & $2.0 $ & $-1.0              $ \\
    $4$  & $200 $ & $0.35 $              & $23$  & $3.56$ & $3.4{\times}10^{-3}$ & $1.0$ & $ 1.0{\times}10^{-1}$ & $4.0$ & $0  $              & $6.3 $ & $ 0               $ \\
    $5$  & $50 $ & $0.625 $              & $13$  & $2.51$ & $2.8{\times}10^{-3}$ & $1.0$ & $ 1.0{\times}10^{-1}$ & $1.5$ & $0  $              & $2.5 $ & $ 0               $ \\
    \hline \hline\\
\end{tabular}

\end{table*}

\begin{table}[!tb]  
  \caption{\label{tab:FitAll2}The parameter values of the MAB model for the anharmonic mode-mode coupling fitted to reproduce the 2D vibrational spectra of liquid water obtained from the full MD simulations.} 
\begin{tabular}{ccccc}
    \hline \hline
    $s$-$s'$   & $3 \tilde{g}_{s^2s'}$  & $3 \tilde{g}_{s{s'}^2}$& $\tilde{\mu}_{ss'}$   & $\tilde{\Pi}_{ss'}$  \\
    \hline    
    $4$-$5$    & $ 0.06              $  & $ 0.06              $  & $-0.05$  & $0.04              $ \\
    $3$-$5$    & $ 0.04              $  & $ 0.04              $  & $0.02$  & $0.01              $ \\
    $3$-$4$    & $ 0.1              $  & $ 0.1              $  & $0.02$  & $0.01              $ \\
    \hline \hline
\end{tabular}
\end{table}  

The present software, originally developed for simulating 2D correlation IR spectra for (1) symmetric stretching, (1') antisymmetric stretching, and (2) bending modes, as intramolecular vibrations of water,\cite{HT25JCP1} has been refined through updates to the integration routines and the discretization scheme of the WDF. 
Because the intramolecular results have been previously reported, here we demonstrate our software focus to the intermolecular vibrational dynamics.
The intermolecular modes of water have been simulated and analyzed using MD.\cite{Saito1995water3modes,Saito1997water3modes,YagasakiSaitoJCP2011Relax,Yagasaki_ARPC64}  Here, we construct the model using the following three modes: The (3) hindered rotational (HR) motion, with 600cm$^{-1}$.\cite{TanakaOhmine1987,ChoOhmine1994} The (4) high-frequency translational (HT) motion or H$_2$O...HOH hydrogen bond stretching mode, with 200cm$^{-1}$, and (5) low-frequency translational (LT) motion or HOH..OH$_2$ hydrogen bond bending mode, with 60cm$^{-1}$.\cite{WalrafenRamanLow,GARBEROGLIO200219}

In the framework of open quantum dynamics, the high-temperature limit ($\beta \hbar \omega \le 1$) coincides with the classical regime. For intermolecular vibrations, even the highest-frequency HR mode (600 cm$^{-1}$) yields $\beta \hbar \omega = 0.45$, validating the use of the CHFPE for accurate description.\cite{T06JPSJ,T20JCP} 

In the past, 2D THz-Raman calculations based on CHFPE were performed for single- and two-mode models.\cite{IIT15JCP,IT16JCP} 
These model parameters were chosen to reproduce the Raman and IR spectra obtained from full MD simulations.\cite{IHT14JCP} 
However, the Raman polarizability function used in MD did not account for intermolecular charge transfer effects: These results are 
inaccurate for describing the Raman signal.\cite{IHT16JPCL,hamm2014} 
Therefore, in this study, we not only adopted the three-mode model but also selected BO model parameters that reproduce the 1D and 2D signals obtained from MD using the newly developed polarization function.\cite{IHT16JPCL} 

The parameter values characterizing anharmonicity of potentials, mode–mode coupling, and the nonlinear components of the dipole moment and/or polarizability were optimized to ensure that the computed 2D THz–Raman spectra faithfully reproduced the peak positions, intensities, and 2D peak profiles observed in the 2D THz–Raman signals obtained from MD simulations, including the effects of intermolecular charge flow.\cite{IHT16JPCL} Currently, such fittings are performed manually; however, machine learning algorithms capable of extracting these parameters directly from MD trajectories are under active development.\cite{UT20JCTC,PJT25JCP1}

To enable direct comparison with previous studies,\cite{IHT14JCP,TT23JCP1,TT23JCP2} we adopted the same formatting conventions as those used in Refs. \onlinecite{HT25JCP1}. Accordingly, the scaling of quantities follows $\tilde{\zeta_s} \equiv \zeta_s (\omega_0/\omega_s)^2$, with $\omega_0 = 4000\,\mathrm{cm}^{-1}$, and $\gamma_s$ is reported as $\gamma_s/\omega_0$. For mode-mode coupling, the parameters are normalized as $\tilde g_{s^3}={g}_{s^3}(\omega_0/\omega_s)^3$, $\tilde g_{s's}={g}_{s's}(\omega_0/\omega_s)(\omega_0/\omega_{s'})$, $\tilde g_{s^2 s'}={g}_{s^2 s'}(\omega_0/\omega_s)^2(\omega_0/\omega_{s'})$, and $\tilde g_{s s'^2}={g}_{s s'^2}(\omega_0/\omega_s)(\omega_0/\omega_{s'})^2$. The fitted system and bath parameter values for the 1D and 2D spectra of liquid water are listed in Tables~\ref{tab:FitAll1} and \ref{tab:FitAll2}, with $\omega_{0} = 4000$ cm$^{-1}$ and $T = 300$ K. 

Note that the 2D THz-Raman spectrum is inherently a convolution of the response function Eq. \eqref{2Dresponse} and the apparatus function, and the response function itself cannot be directly extracted.\cite{HammTHz2012,Hamm2013PNAS,hamm2014,Hamm20252DTHz,Blake20162DThzRaman,Blake20172DThzRaman,Blake20192DThzRaman,Blake20202DThzRaman}
It is known that MD-based response functions fail to reproduce experimental results, and no existing theory currently accounts for the experimental results. Although the model parameters used here were derived from MD simulations for 2D THz-Raman,\cite{IHT16JPCL} they alone are insufficient for reproducing the observed spectra. 
Nevertheless, the present BO model offers flexibility in tuning parameters such as anharmonicity and polarization. Through iterative adjustment, it is possible to reproduce experimental features and subsequently analyze the data.  However, this extension is beyond the scope of this demonstration and will be pursued in future work.

The 1D IR and Raman response functions, expressed as $R_{\mathrm{IR}}(t)$ and $R_{\mathrm{Raman}}(t)$ are defined by Eq. \eqref{1Dresponse} with $A(\bm{q})=\mu(\bm{q}) $ and $A(\bm{q})={\Pi} (\bm{q})$, respectively. Depending on how Raman pulses are applied, the 2D THz-Raman responses are classified into three types: 
(i) $R_{\mathrm{RTT}}^{(3)} (t_{2},t_{1})$ for Raman-THz-THz (RTT), (ii) $R_{\mathrm{TRT}}^{(3)} (t_{2},t_{1})$ for THz-Raman-THz (TRT), and (iii) $ R_{\mathrm{TTR}}^{(3)} (t_{2},t_{1})$ for THz-THz-Raman (TTR), respectibely, defined by Eq.\eqref{2Dresponse} with (i) $A(\bm{q})={\Pi} (\bm{q})$, (ii) $B(\bm{q})={\Pi} (\bm{q})$, and (iii) $C(\bm{q})={\Pi} (\bm{q})$ and the remaining operator in each case is ${\mu} (\bm{q})$.
 In the HEOM formalism, the Green's function $\mathcal{G}(t)$ is the time evolution of the CHFPE, and $W^{\mathrm{eq}}(\bm{q}, \bm{p})$  is the equilibrium
distribution, which is described in terms of CHFPE elements.\cite{IT16JCP,T06JPSJ,T20JCP,TS20JPSJ,KT04JCP,HT25JCP1,T98CP}

\begin{figure}
\centering
\includegraphics[width=80mm]{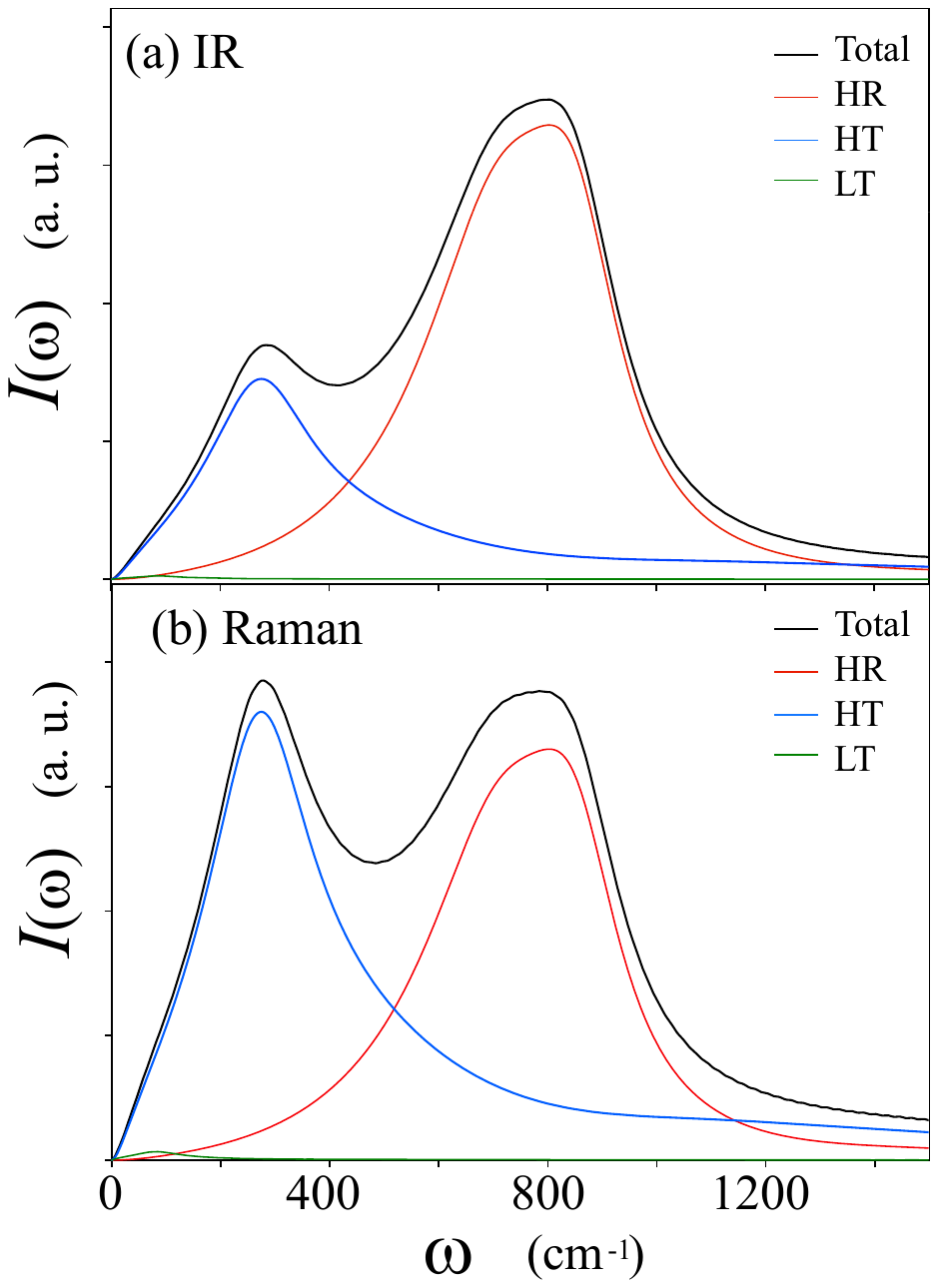}
\caption{\label{1DTHzRaman}
{The computational results of (a) THz and (b) parallelly polarized (VV) Raman.
In each figure, the red, blue, and green curves represent the contributions from hindered rotation (HR), translational (HT), and translational (LT) motion, respectively.
The spectral intensities are normalized with respect to the absolute value of the peak intensities of the total contribution.}}
\end{figure}

Numerical calculations were performed on two PCs: (i) Intel(R) Core(TM) i9-9900K 8-core, RAM: 32GB, GPU: GeForce RTX 3080 Ti 12GB, and (ii) CPU: Intel(R) Xeon(R) Gold 6212U, RAM: 192GB, GPU: A100 GPU 40GB. The OS we used was almalinux8, the compiler was gcc 8.5.0, and we used CUDA 12.6 and Python 3.12. The libraries used in C++/CUDA are hdf5-1.14.6, gputt, cudart, cublas, cufft. The libraries used in Python are Optuna, numpy, matplotlib, h5py, and pybind11. The time required for the calculations (equivlation $\Rightarrow$  1DIR $\Rightarrow$  1DRaman $\Rightarrow$ 2DRTT $\Rightarrow$  2DIRI $\Rightarrow$  2DIIR) was approximately two hours for both (i) and (ii). Note that while (ii) can run three jobs simultaneously, but it is about 10\% slower than (i).

Figure ~\ref{1DTHzRaman} shows (a) THz and (b) parallelly polarized (VV) Raman spectra obtained from the three-mode models (black).  
The 1D spectra were obtained using the parameter values listed in Tables \ref{tab:FitAll1} and \ref{tab:FitAll2}. Note that the 1D spectra profiles do not change when the mode-mode coupling is not included because the contribution of each mode to the spectral intensities is significantly stronger than the contribution of the mode-mode coupling. To elucidate the contribution of the hindered rotational mode, we decomposed the 1D spectra into (3) HR (red curve), (4) HT (blue curve), and (5) LT (green curve) motions.

\begin{figure*}[!t]
\centering
  \includegraphics[scale=0.9]{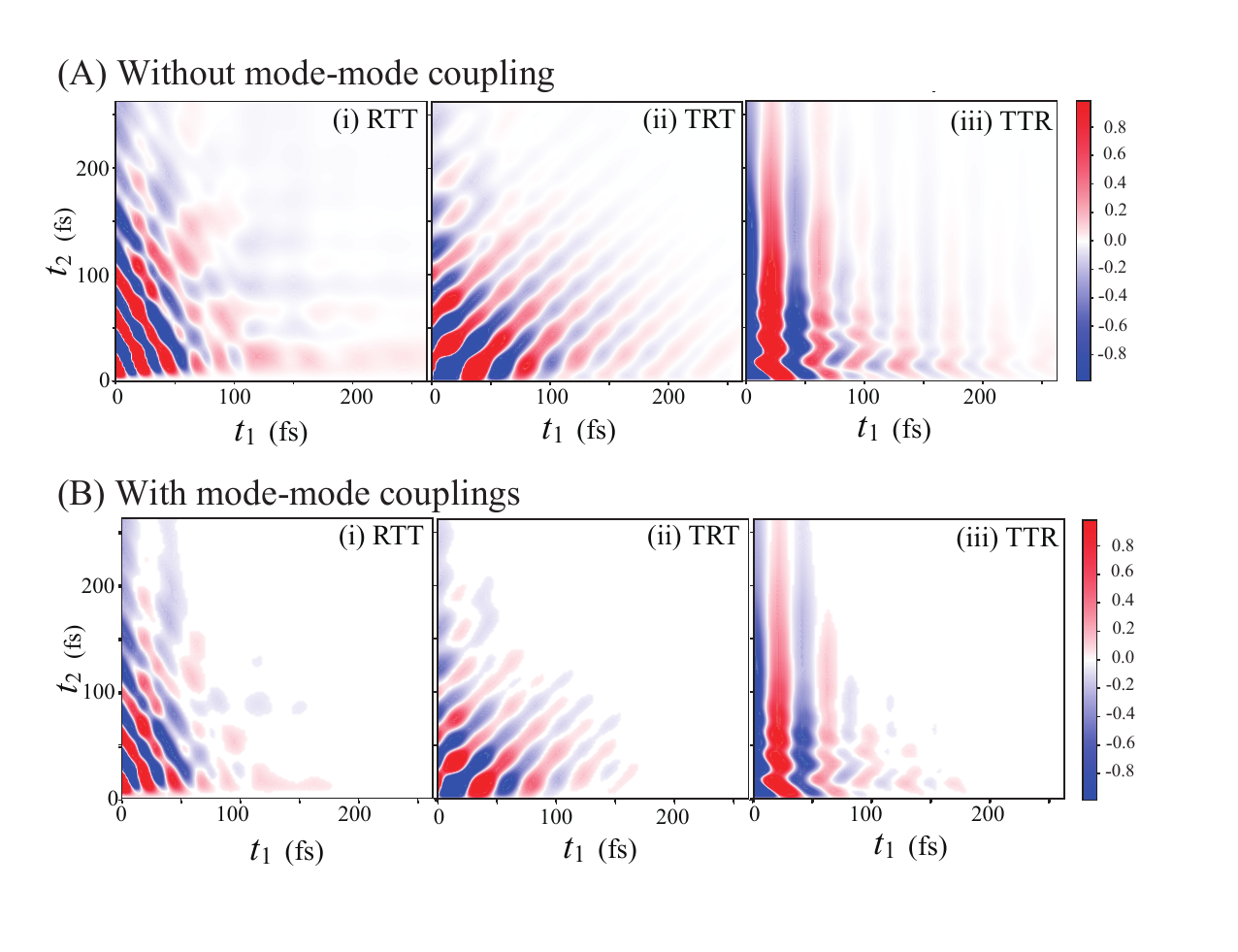}
\caption{\label{2DTHzRaman}
The 2D THz-Raman signals of water were calculated using the three-mode model in two cases: (A) when all mode-mode coupling was set to zero, and (B) in the case listed in Table \ref{tab:FitAll2}. For each case, (i) the 2D Raman-THz-THz (RTT), (ii) 2D THz-Raman-THz (TRT), and (iii) 2D THz-THz-Raman (TTR) signals were depicted. 
Red and blue shadings represent positive and negative signals, respectively. Signal intensities are normalized to the absolute value of the peak amplitude. The relative intensities are color-coded, with positive values shown in red and negative values in blue, as indicated by the color bar.
}
\end{figure*}

Figure ~\ref{2DTHzRaman} shows the (i) 2D RTT, (ii) 2D TRT, and (iii) 2D TTR  calculated from the three-mode model. To investigate the effect of mode-mode coupling on 2D profiles, we performed calculations for two cases: (A) when all mode-mode coupling is zero, and (B) the cases listed in Table \ref{tab:FitAll2}. In the 1D IR and 1D Raman spectra presented in Fig. \ref{1DTHzRaman}, there is no change in the profile when mode-mode coupling is included, but in the 2DTHz-Raman, we see a suppression of the elongated peak with respect to time due to energy relaxation between modes.

As can be seen in comparison with the full MD results,\cite{IHT16JPCL} the three-mode model reproduces the qualitative features of the peak profiles in the 2D spectra.  Compared to one-mode\cite{IHT14JCP} and two-mode BO calculations,\cite{IT16JCP} the three-mode model is more descriptive of 2D spectral profiles.

This indicates that three-mode model analysis captures essential intermolecular features.
The role of individual terms in the total Hamiltonian [see Eq. \eqref{sec:Total Hamiltonian}] in the context of the 2D spectral profiles have been studied in detail, depending on their parameter values.\cite{IIT15JCP,IT16JCP} 
Using the extracted model parameter values, we can analyze the effects of anharmonicity, anharmonic mode-mode coupling, nonlinear polarizability and dipole moment, relaxation phenomena, and vibrational dephasing processes for each mode.\cite{IIT15JCP,IT16JCP}  This approach remains viable even when the 2D spectral peaks derived from MD simulations are overlapping or indistinct. 
While the present calculations for 2D THz-Raman signals serve as a demonstration for running the software; a separate paper will present a detailed quantitative analysis of the experimental results in the frequency domain.\cite{HammTHz2012,Hamm2013PNAS,hamm2014,Hamm20252DTHz,Blake20162DThzRaman,Blake20172DThzRaman,Blake20192DThzRaman,Blake20202DThzRaman}

\section{Conclusion}
\label{sec:conclude} 

The environment, encompassing solvents, proteins, and solid matrices, is pivotal in driving chemical processes, including reactions and energy transfers, by mediating thermal fluctuations and dissipation.
Ultrafast nonlinear laser spectroscopy, especially coherent multidimensional spectroscopy, is a key experimental method for studying these effects due to its sensitivity and versatility.\cite{TM93JCP,mukamel1999principles,Cho2009,Hamm2011ConceptsAM,HammPerspH2O2017} 
Theoretical support is crucial for analyzing complex 2D spectral profiles, as they capture environmental details but are highly sensitive to approximations and numerical errors. This presents challenges in non-equilibrium thermodynamics and statistical mechanics.

This study introduces a flexible simulation tool for qualitative analysis of 2D spectral profiles.
Although the theory relies on a model, it can accurately simulate spectral profiles, capturing effects like thermal fluctuations, dissipation, and vibrational relaxation that lead to inhomogeneous broadening. 

The MD results play a key role in determining modeling parameters,\cite{UT20JCTC} while the present findings complement them by enabling analysis of spectral profiles observed in MD simulations. Quantum effects play a pivotal role in intramolecular vibrational dynamics. Capturing these effects requires trajectories obtained via quantum MD.\cite{JianLiu2018H2OMP,Imoto_JCP135} To rigorously address quantum behavior, future developments will incorporate the QHFPE, which enables a quantum-mechanically ``exact'' solution of the same Hamiltonian for CHFPE.\cite{TT23JCP1,TT23JCP2} Although current implementations of QHFPE are restricted to two vibrational modes, extension to three modes is essential to achieve consistency with the present results. Quantum calculations of 2D spectra,\cite{ST11JPCA,TT23JCP1,TT23JCP2} including reactive processes such as proton transfer\cite{IT05JCP,ZBT21JCP} and nonadiabatic transitions,\cite{IT18CP,IT19JCTC,IDT19JCP} are becoming increasingly feasible through BO-based approaches, thereby addressing limitations inherent in MD. Importantly, the spectral calculation module remains modular and reusable, positioning the present program as a versatile computational platform for both classical and quantum 2D spectroscopy.

While the present demonstration focuses on water, the theoretical framework is general and applicable to other solutions. 
The accompanying source code is designed to accommodate diverse systems by modifying the potential energy functions, mode-mode couplings, and system–bath interactions with minimal effort. Owing to the classical nature of intermolecular vibrational modes, the present implementation serves as a versatile platform for simulating 2D THz-Raman and Raman spectroscopies across a broad range of condensed-phase environments.

Finally, we note a clear similarity between the particle system interacting with an anisotropic heat bath in 3D space\cite{YKT25JCP1,ZT25JCP1} and the three-mode system discussed here. This resemblance suggests that the current program can be adapted to simulate and analyze 2D spectra of a quantum dissipative system in 3D space.

\section*{Supplementary Material}
Numerical integration codes for 1D IR, 1D Raman,  2D THz-Raman, 2D THz-IR, and 2D IR are provided as supplemental materials. The manual can be found in the ReadMe.pdf file.

\section*{Acknowledgments}
Y. T. was supported by JST (Grant No. CREST 1002405000170).  

\section*{Author declarations}
\subsection*{Conflict of Interest}
The authors have no conflicts to disclose.

\section*{Data availability}
The data that support the findings of this study are available from the corresponding author upon reasonable request.

\appendix

\section{The CHFPE in the Hermite polynomial representation}
\label{Sec:HEOMDIS}

The equations of motion for $c^{(\bm{n})}_{k, l,m}({\boldsymbol q};t)$ are then expressed as\cite{IIT15JCP,IT16JCP}
\begin{eqnarray}
&&\frac{\partial c^{(\bm{n})}_{k, l,m}({\boldsymbol q};t)}{\partial t} = \nonumber \\
&&\sqrt{k+1}\hat{D}^{+}_{1}c^{(\bm{n})}_{k+1, l,m}({\boldsymbol q};t)
- \sqrt{k}\hat{D}^{-}_{1}c^{(\bm{n})}_{k-1, l,m}({\boldsymbol q};t) \nonumber \\
&+& \sqrt{l+1}\hat{D}^{+}_{2}c^{(\bm{n})}_{k, l+1,m}({\boldsymbol q};t)
- \sqrt{l}\hat{D}^{-}_{2}c^{(\bm{n})}_{k, l-1,m}({\boldsymbol q};t) \nonumber \\
&+& \sqrt{m+1}\hat{D}^{+}_{3}c^{(\bm{n})}_{k, l,m+1}({\boldsymbol q};t)
- \sqrt{m}\hat{D}^{-}_{3}c^{(\bm{n})}_{k, l,m-1}({\boldsymbol q};t) \nonumber \\
&-& n_1\gamma_{1}c^{(\bm{n})}_{k, l,m}({\boldsymbol q};t)
- n_2\gamma_{2 }c^{(\bm{n})}_{k, l,m}({\boldsymbol q};t) 
- n_3\gamma_{3 }c^{(\bm{n})}_{k, l,m}({\boldsymbol q};t) \nonumber \\
&-& n_1\gamma_1{\Theta}^{(1)}_{k+1}c^{(\bm{n}-\bm{e}^{(1)})}_{k+1, l,m}({\boldsymbol q};t)
      - {\Phi}^{(1)}_{k-1}c^{(\bm{n}+\bm{e}^{(1)})}_{k-1, l,m}({\boldsymbol q};t) \nonumber \\
&-& n_2\gamma_2{\Theta}^{(2)}_{l+1}c^{(\bm{n}-\bm{e}^{(2)})}_{k, l+1,m}({\boldsymbol q};t)
- {\Phi}^{(2)}_{l-1}c^{(\bm{n}+\bm{e}^{(2)})}_{k, l-1,m}({\boldsymbol q};t)  \nonumber \\
&-& n_3\gamma_3{\Theta}^{(3)}_{m+1}c^{(\bm{n}-\bm{e}^{(3)})}_{k, l,m+1}({\boldsymbol q};t) 
- {\Phi}^{(3)}_{m-1}c^{(\bm{n}+\bm{e}^{(3)})}_{k, l,m-1}({\boldsymbol q};t), \nonumber \\
\label{HEOM2FP-H}
\end{eqnarray}
where
\begin{eqnarray}
\hat{D}^{\pm}_{s} &=& \frac{1}{2}\frac{1}{\sqrt{m_sk_BT}}\frac{\partial U({\boldsymbol q})}{\partial q_s} {\mp} \frac{1}{m_s}\sqrt{{m_sk_BT}}\frac{\partial}{\partial q_s}, \\
{\Theta}^{(s)}_{j} &=& -\zeta_s\frac{V^{(s)}({\boldsymbol q}_{s})}{\partial q_s}\sqrt{{m_s}k_BT}\sqrt{j}, 
\end{eqnarray}
and 
\begin{eqnarray}
{\Phi}^{(s)}_{j} &=& 
\frac{V^{(s)}({\boldsymbol q}_{s})}{\partial q_s}\frac{1}{\sqrt{m_{s}k_BT}}\sqrt{j+1}.
\end{eqnarray}
In the numerical calculations, we employed the dimensionless coordinate $\overline{q}_s$.

\bibliography{tanimura_publist,HT24,TT23}

\end{document}